\documentclass[%
reprint,
amsmath,amssymb,
aps
]{revtex4-1}

\usepackage[normalem]{ulem}
\usepackage{soul}

\usepackage{graphicx}
\usepackage{dcolumn}
\usepackage{xcolor}

\begin{document}

\title{On the relative role of the physical mechanisms \\ on complex biodamage induced by carbon irradiation
\\
\medskip
}

\author{Simone Taioli$^{a,b,c}$}
\email{taioli@ectstar.eu}
\author{Paolo E. Trevisanutto$^{a,b,d}$} 
\author{Pablo de Vera$^e$}
\author{Stefano Simonucci$^{f,g}$}
\author{Isabel Abril$^h$}
\author{Rafael Garcia-Molina$^e$}
\author{Maurizio Dapor$^{a,b}$}
\email{dapor@ectstar.eu}

\affiliation{$^a$European Centre for Theoretical Studies in Nuclear Physics and Related Areas (ECT*-FBK)}
\affiliation{$^b$Trento Institute for Fundamental Physics and Applications (TIFPA-INFN), Trento, Italy}
\affiliation{$^c$Peter the Great St. Petersburg Polytechnic University, Russia}
\affiliation{$^d$Center for Information Technology, Bruno Kessler Foundation, Trento, Italy}
\affiliation{$^e$Departamento de F{\'i}sica, Centro de Investigaci{\'o}n en {\'O}ptica y Nanof{\'i}sica, Universidad de Murcia, Spain}
\affiliation{$^f$School of Science and Technology, University of Camerino, Italy}
\affiliation{$^g$INFN, Sezione di Perugia, Italy}
\affiliation{$^h$Departament de F{\'i}sica Aplicada, Universitat d'Alacant, Spain}

\begin{abstract}
The effective use of swift ion beams in cancer treatment (known as hadrontherapy) as well as an appropriate protection in manned space missions rely on the accurate understanding of energy delivery to cells damaging their genetic information.
The key ingredient characterizing the response of a medium to the perturbation induced by charged particles is its electronic excitation spectrum.
By using linear response time-dependent density functional theory, we obtain the energy and momentum transfer excitation spectrum (the energy-loss function, ELF) of liquid water (main constituent of biological tissues), in excellent agreement with experimental data. The inelastic scattering cross sections obtained from this ELF, together with the elastic scattering cross sections derived considering the condensed phase nature of the medium, are used to perform accurate Monte Carlo simulations of the energy deposited by swift carbon ions in liquid water and carried away by the generated secondary electrons producing inelastic events (ionization, excitation, and dissociation electron attachment, DEA), strongly correlated with cellular death, which are scored in sensitive volumes having the size of two DNA convolutions. The sizes of clusters of damaging events for a wide range of carbon ion energies, from those relevant to hadrontherapy up to cosmic radiation, predict with unprecedented statistical accuracy the nature and relative magnitude of the main inelastic processes contributing to radiation biodamage, confirming that ionization accounts for the vast majority of complex damage.
DEA, typically regarded as a very relevant biodamage mechanism, surprisingly plays a minor role in carbon-ion induced clusters of harmful events.  

\end{abstract}



\maketitle

The interaction of swift ions with matter is used  to probe its structure, to characterize and modify materials properties or to design microdevices 
\cite{Nastasi1996}.

Recently, energetic ion beams have also found application in treating cancer (hadrontherapy), owing to their peculiar energy delivery to the target material, characterized by a depth-dose profile with a sharp peak (so called Bragg peak) at the end of their range, where they produce severe cell damage, while sparing effects on both traversed and deeper located healthy tissues \cite{Nikjoo-Uehara-Emfietzoglou2012,Ma-Lomax2013,Surdutovich2014}. The higher spatial resolution of the dose delivered by ion beams as compared to conventional radiation (photons and electrons), is not their only advantage. The capacity of carbon ions to produce complex DNA lesions, which eventually lead to cell killing, is very promising for developing more effective oncological treatments \cite{Suit2009,Amaldi2010} when radiation damage to surrounding normal tissue could be dangerous or even deadly \cite{Ebner2016}. 

On the other hand, highly energetic ions with high atomic number are also abundant in cosmic radiation. This fact poses risks to safety of humans and devices on board of space stations, representing a serious challenge on space exploration  
\cite{Durante-Cucinotta2011,VignoliMuniz2017}.
Even the required coherence  in quantum computers can be destroyed by cosmic rays \cite{Vepsalainen2020}.




Achieving optimal results for selective cell destruction requires understanding of the fundamental processes that produce most harmful lesions occurring at cellular level. Among these, clustering of sugar-phosphate strand and base pair breaks in the DNA structure are mostly lethal for cell survival 
\cite{Goodhead1994,Alizadeh2015,Verkhovtsev2016}.
Such clustered DNA lesions might be induced by several physical and chemical mechanisms \cite{Mozumder2004,Solovyov2017}. As for the primary physical events, these are mainly produced by the secondary electrons abundantly generated by the incident ions. In particular, the latter can induce electronic excitations and ionizations, as well as dissociative electron attachment (DEA) at energies of a few eV, leading to molecular fragmentation. Since most of the electrons are ejected with energies below 50 eV, DEA has been regarded as  especially relevant to biodamage \cite{Boudaiffa2000}.

In contrast, current physical approaches to measure clustered inelastic events in targets with DNA size, referred to as nanodosimetry \cite{Conte2017, Conte2018} are based on the exclusive measurement of ionizing collisions (in gas-phase detectors). In this context, it is crucial to precisely determine the relative contribution to clustered DNA damage for carbon ions due to 
the different interaction mechanisms. This problem can be carefully addressed by means of track-structure Monte Carlo (MC) simulations, provided that reliable cross sections of the different interaction mechanisms with the target medium are used \cite{Nikjoo-Uehara-Emfietzoglou2012}. Typically, radiation transport is studied in liquid water, which is conventionally considered as a good proxy for biological tissue, being its main constituent.

In this work we accurately describe the liquid water electronic excitation spectrum from first principles and  simulate in detail, with a remarkable high 
statistics, the processes occurring around the swift carbon ion track. The relative role of ionizations, dissociative electronic excitations and DEA to the clustering  of damaging events in the medium  is quantified for the first time, from typical ion energies around the Bragg peak --useful in hadrontherapy-- to cosmic rays energies --relevant to space missions.  

\section*{Results and discussion}
\label{ssubsec:ELF}
The interaction of charged particles through a medium depends mainly on its electronic excitation spectrum, encoded in the energy loss function ELF$(k,E)$ \cite{Lindhard1954,Ritchie1959}, where $\hbar k$ and $E$ are the momentum and energy transfer, respectively. The ELF, conveniently weighted and integrated, provides the  probabilities for the most important inelastic scattering processes occurring at the passage of charged particles inside a condensed medium \cite{Nikjoo-Uehara-Emfietzoglou2012}. 
Liquid water is the most abundant constituent of living beings, thus a great deal of effort has been put to obtain its ELF over the Bethe surface $(\hbar k,E)$. Unfortunately, this quantity is only known experimentally for a limited range of excitation energies $E$ and for a finite set of momentum transfers $\hbar k$ \cite{Watanabe1997,Hayashi2000,Watanabe2000}. This lack of data hinders the study of energetic particles moving through biological media. Theoretical estimates could fill the gap of required data. However, \textit{ab initio} computations, being mostly carried out for water in gas phase instead of liquid --i.e. condensed-- state that appears in living tissues, typically do not compare well with the available experimental data over the entire Bethe space. Moreover, first-principles simulations are cumbersome owing to their high computational cost, particularly when dealing with systems characterized by random molecular arrangements, such as liquid water. In principle, the computational complexity of calculating the ELF of liquid water by time-dependent density functional theory (TDDFT) for all $\hbar k$ and $E$ (see Methods section for details) is largely due both to the size of the supercell necessary to describe its non-crystalline structure and to the need of simulating a number of different molecular configurations to achieve statistical significance of the averaged final result. However, we have discovered how to limit the calculations to one particular cell configuration (32 water molecules at the experimental density in room conditions, $\rho=1$ g/cm$^3$), which leads to a considerable reduction of the computational effort, limiting the ELF calculation to one particular snapshot. Figure 1 depicts the calculated ELF of liquid water compared to available experimental data \cite{Watanabe1997,Hayashi2000,Watanabe2000}, showing an excellent agreement in a wide range of energy and momentum transfers.
\begin{figure}[t]
\centering
\includegraphics[width=\linewidth]{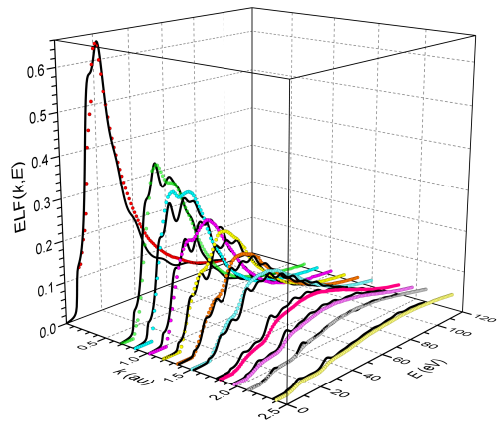}
\caption{Energy loss function (ELF) of liquid water as a function of energy $E$ and momentum $\hbar k$ transfer. Black continuous curves represent TDDFT calculations, whereas symbols report experimental data \cite{Watanabe1997, Hayashi2000, Watanabe2000}. 
}
\label{fig:ELF}
\end{figure}

\label{ssubsec:ElectronGeneration}
The passage of swift charged particles through a medium prompts the generation of secondary electrons (SE), which leads to subsequent cascade processes where these secondaries produce further electrons. All these charged particles lose energy and change their direction of motion by colliding with the medium constituents. It is currently admitted that the collision of SE with the DNA from biological tissues can trigger irreparable damage if critical events, such as bond breaking of the nucleobase pairs or of the sugar-phosphate chain, occur. The biodamage effectiveness is enhanced when these episodes happen close enough (i.e., clustering) in the DNA molecule, as this event hampers the repair mechanisms \cite{Nikjoo2016RepProgPhys}. The occurrence of these damaging events strongly depends on the energy with which the electrons reach the DNA molecule. Therefore, for better characterizing the physical stage in radiation biodamage it is crucial to know most accurately the characteristic energy range of the electrons reaching typical DNA volumes.
  
The main quantities to study the generation, propagation and effects of the SE produced by energetic ion beams in condensed phase media are the probability of ionization and the energy and angular distributions of the emitted electrons. These quantities are encoded in the cross sections, which can be accurately calculated using the dielectric formalism \cite{Lindhard1954,Ritchie1959,Nikjoo-Uehara-Emfietzoglou2012} once the projectile (mass, charge and energy) and medium (electronic excitation spectrum) characteristics are known (see Methods section for details).
  


\begin{figure}[t]
\centering
\includegraphics[width = \linewidth]{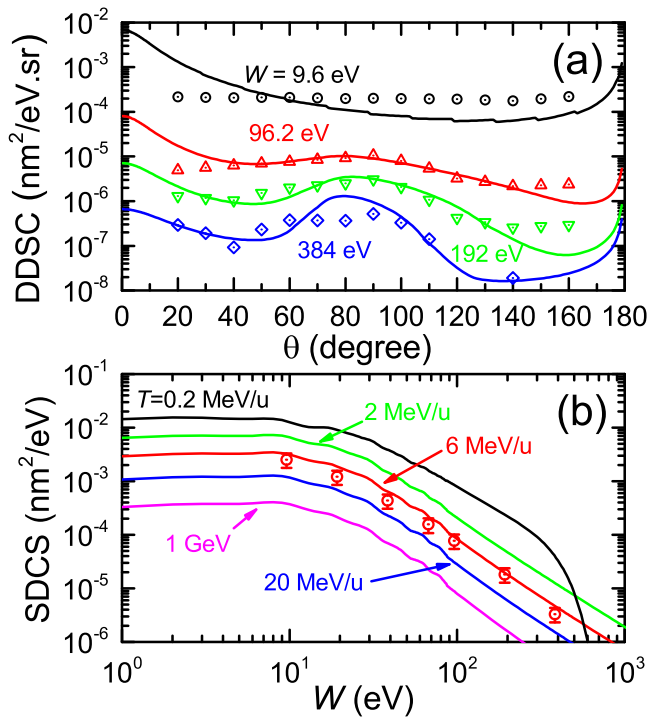}
\caption{(a) Angular distribution of electrons with energy $W$, generated by $T=6$ MeV/u C ions in liquid water. (b) Energy distribution of electrons, generated in liquid water by C ions with initial kinetic energy $T$. Continuous lines represent our  calculations, whereas symbols report experimental data in water vapor \cite{DalCappello2009}. 
} 	
\label{fig:DDCS-W-theta}
\end{figure}

The doubly differential cross section (DDCS) for the angular distribution of electrons ejected with energy $W$, per unit solid angle d$\Omega$, can be obtained from the ELF \cite{deVera2015}, using the procedure discussed in the Methods section. This DDCS is depicted in Fig. \ref{fig:DDCS-W-theta}(a) together with available experimental measurements in the gas phase for carbon ions of kinetic energy (KE) $T=6$ MeV/u \cite{DalCappello2009}. The agreement with the experimental data is fairly good for the wide range of electron energies $W$ and angles $\theta$, despite the phase differences between calculations and experiments. It is worth noting the rather isotropic angular distribution of the lower energy electrons.

Appropriate integration of the electronic excitation spectrum for all possible values of the momentum transfer $\hbar k$ provides the singly differential cross section (SDCS) of emission of an electron with energy $W$ (see Methods). Figure \ref{fig:DDCS-W-theta}(b) compares the calculated SDCS with available experimental data for water vapor, for several incident carbon ion KE $T$, from the typical values at the Bragg peak in hadrontherapy ($0.2$ MeV/u), to the very high ones appearing in cosmic radiation (1 GeV = 83.33 MeV/u). Comparison with measurements at $T=6$ MeV/u \cite{DalCappello2009} shows a good accordance, within the experimental error bars. SE energy distributions are characterized by a peak at $W \simeq 10$ eV, rapidly decreasing at higher energy. The SDCSs drop to zero at the maximum energy at which SE electrons can be ejected. It is remarkable the abundance of low-energy electrons, to which it is attributed a significant role in clustered damage on the nanometer scale \cite{Boudaiffa2000,Alizadeh2015,Rezaee2014}.


\label{ssubsec:ElectronPropagation}

The energy transferred by the carbon ions to SE is carried away from the ion path as they propagate through the medium, where they undergo elastic collisions (leading to trajectory deviation), and inelastic interactions (resulting in electronic excitations, further ionizations, electron-phonon coupling, and trapping phenomena).

\begin{figure}[t]
	\centering
	\includegraphics[width = \linewidth]{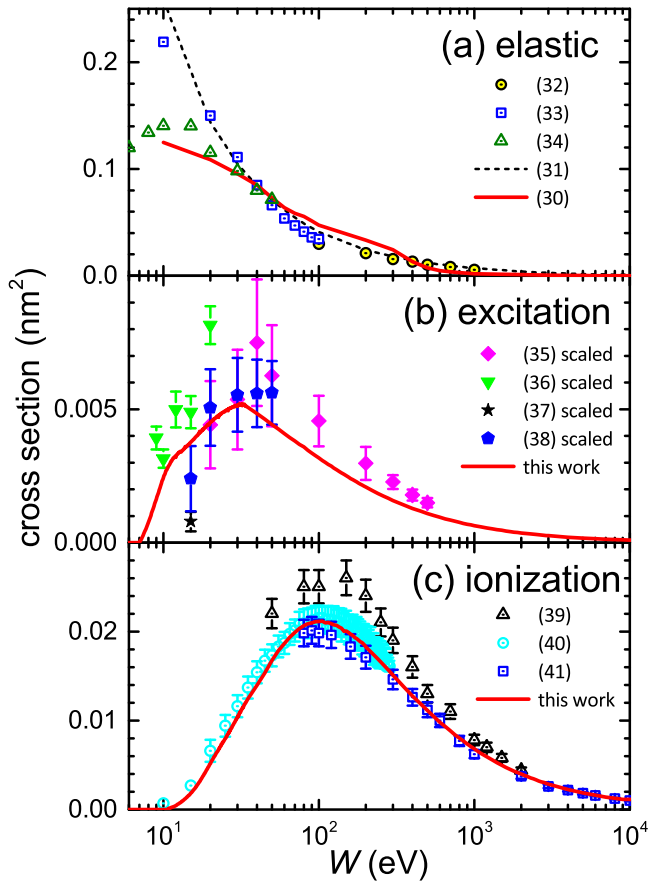}
	\caption{
	Cross sections of electrons in liquid water as a function of their energy $W$ due to (a) elastic, (b) excitation and (c) ionization processes. Curves correspond to calculations as explained in the text.
	Symbols are experimental data for water vapor: (a) \cite{Itikawa2005, Katase1986,Cho2004}, (b) \cite{Thorn2007, Ralphs2013, Brunger2008, Matsui2016} (scaled) and (c) \cite{Bolorizadeh1986, Bull2014, Schutten1966}. 
}
\label{fig:eIMFP}	
\end{figure}

Electron elastic scattering is reckoned by directly solving the Dirac-Hartree-Fock equation for a cluster of six water molecules, to account for multiple scattering from surrounding water molecules in liquid phase, using a projected potential approach \cite{Taioli2010,taioli2009surprises} based on Gaussian functions \cite{morresi2018nuclear,Taioli2010,taioli2009surprises} (see Methods section for details). In Fig. \ref{fig:eIMFP}(a) we compare the elastic cross section we have obtained using the water cluster (red curve) and the widely adopted Mott cross section \cite{Mott1929} (dashed line). The latter is in excellent agreement with the recommended experimental data for water vapor (blue squares in Fig. \ref{fig:eIMFP}(a))  \cite{Katase1986,Itikawa2005}. Notice that the single molecule elastic cross section calculation obtained by using our Gaussian-based relativistic projected-potential approach provides almost identical results to the Mott cross section. At odds, the condensed phase nature of liquid water emerges as a significant deviation from the single water molecule case (typically used in hadrontherapy modeling), particularly at low energy where the elastic cross section assessed on the cluster is appreciably reduced with respect to the single molecule, a behavior that also appears in the experimental data by Cho \textit{et al}. \cite{Cho2004}.

The main inelastic processes of electron projectiles (ionization and excitation of target electrons) are dealt with the dielectric formalism (see Methods section) by replacing the ion characteristics with the electron ones and by introducing indistinguishability and exchange. Upon appropriate integration of the ELF for all $\hbar k$ and $E$ transfers and by properly accounting for ionization or excitation events \cite{deVera2019}, the relevant cross sections are obtained (see Figs. \ref{fig:eIMFP}(b) and \ref{fig:eIMFP}(c) for excitation and ionization, respectively).
Our calculated electronic excitation and ionization cross sections are compared, respectively, with experimental values available for a few specific excitation channels in gas phase \cite{Thorn2007, Brunger2008, Ralphs2013, Matsui2016} (scaled to include all known channels \cite{Thorn2008}) and for ionizations \cite{Schutten1966,Bolorizadeh1986,Bull2014} in a broad energy range ($10$ - $10^4$ eV). Despite the rather scattered experimental data, the excitation and ionization cross sections obtained from the ELF of liquid water exhibit a general shape and magnitude in fairly good agreement with the experimental results in the entire energy range.
Remarkably, our calculations agree almost perfectly with the most recent experimental data \cite{Bull2014}. 

\begin{figure}[t]
	\centering
	\includegraphics[width=\linewidth]{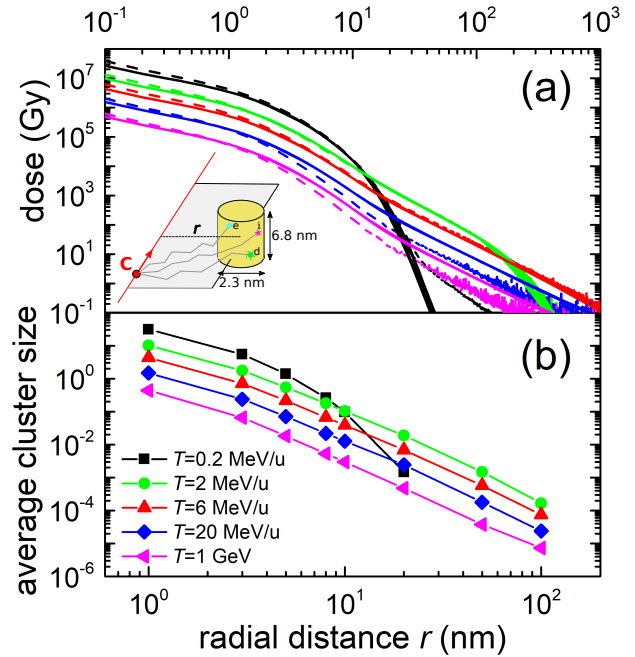}	
	\caption{(a) Dose deposited by carbon ions in liquid water as a function of the distance $r$ from the ion track, for several ion energies $T$. Elastic collisions were accounted for using the Mott cross section for a single water molecule (dashed lines) or the improved treatment taking into account the condensed phase nature of liquid water (solid lines).
	(b) Average cluster size $M_1$ in a sensitive volume of liquid water having the dimension of two DNA turns, as a function of the radial distance $r$ from the ion track, for different values of the carbon ion energy $T$. Symbols correspond to simulations, with lines plotted to guide the eye. The inset depicts a scheme of the nanometric cylinder used in the scoring of damaging events: excitation (e), ionization (i) and DEA (d). 
	}
	\label{fig:Dose}
\end{figure}

Besides the previous inelastic processes, at energy $\lesssim 20$ eV electrons may induce quasi-particle excitations (most notably phonons \cite{Frohlich1954} and polarons \cite{Ganachaud1995}) or also DEA \cite{Boudaiffa2000}.

The electronic cross sections corresponding to different processes are input functions to the SEED MC code \cite{Dapor2017,Dapor2020book}, which simulates the electron transport (and ensuing generation) through liquid water (see Methods). Electrons deposit energy around the carbon ion path, resulting in the radial dose distribution depicted in Fig. \ref{fig:Dose}(a) for several ion energies, when condensed phase effects (i.e., the improved cross section, reported by continuous lines) or only single atoms are considered in elastic collisions (i.e., the widely used Mott cross section \cite{Mott1929}, dashed lines). 
The quality of the elastic scattering cross section has a noticeable impact on the energy deposition distribution in the nanometric length scale, with largest discrepancies at increasing distances from the ion's track at all KEs, especially for the higher ones. For the lowest ion energy studied (0.2 MeV/u, characteristic of the Bragg peak), the radial dose is both the most intense and most concentrated in a 10-20 nm region around the ion track, which corresponds to the maximum distance that the most energetic electrons can travel (as seen from the kinematic cut-off in Fig. \ref{fig:DDCS-W-theta}(b)). As the energy of the ions increases, electrons can be ejected at higher KEs, which translates in more extended radial doses; however, the corresponding decrease of the ionization cross section at higher ion energy lowers the number of emitted electrons, then reducing the radial dose absolute intensity.

The energy deposition in biological tissues can lead to cell damage (with its death or avoidance being the ultimate goal of hadrontherapy and radiation protection, respectively). However, it is an integrated quantity measuring the deposited macroscopic energy and, thus, is not the
most convenient observable to assess biological damage. Indeed, it is the clustering of damaging events in nanometric volumes, mimicking the size of DNA, that determines biological effects, such as lethal lesions. Typically, these consist in clustered bond breaking of DNA molecules, a damage difficult to repair by the cell machinery that leads eventually to cell mutation or death \cite{Goodhead1994}.

Figure \ref{fig:Dose}(b) shows the simulation \cite{Dapor2017, Dapor2020book} of the average cluster size $M_1$ of damaging events (detailed in the next paragraph) scored in nanometric cylinders of liquid water having 2.3 nm-diameter and 6.8 nm-height (the size of 20 base  pairs of DNA), for several carbon ion energies $T$ and impact parameters $0 \le r \le 100 \mbox{ nm}$. To consider the inherently stochastic nature of this process a large number of ion's shots at each ion kinetic energy have been conducted (as explained in Methods), minimizing the statistical uncertainties (error bars in Fig. \ref{fig:Dose}(b) are smaller than the symbol size).

We considered as damaging mechanisms: i) ionizations, ii) electronic excitations leading to bond dissociation, iii) and DEA events, the latter being simulated using the experimental cross section available for water molecules \cite{Itikawa2005,taioli2006wave}. We have assumed that only $40\%$ of electronic excitation events lead to molecular dissociation \cite{Thorn2008} and thus produce damage. $M_1$ follows rather closely the shape of the radial doses. For 0.2 MeV/u ions, average cluster sizes are significantly large ($\gtrsim 10$) at ion-target impact parameters (IP) $r < 3$ nm, being always larger than 1 for $r < 5$ nm. Cluster size drops to 0 at $r > 20$ nm, meaning that SE cannot travel beyond that distance at this ion KE. Cluster size distributions progressively lower at increasing ion energy, although they are still larger than 1 at energies below 6 MeV/u with closer values of $r$ ($\sim 3$-4 nm at 2 MeV/u and $\sim 2$ nm at 6 MeV/u). The highest energy (1 GeV) ions considered in this work are not capable to induce clusters of average size $\ge 1$ for any IP. These trends show the potential of different ion KEs to induce irreparable DNA damage: low KE ions around the Bragg peak region are especially harmful, producing large damage cluster sizes, while individual high energy ions will not produce significant lethal damage.

From the cluster size distributions, it is possible to obtain useful statistical information on the probability of inducing complex DNA damage, which is the necessary information to relate physical damage with biological outcomes.
This is the purpose of experimental nanodosimetry \cite{Conte2018}, which employs gas-phase detectors to estimate the probabilities of clustering of inelastic events (estimated as ionizing collisions) in volumes of dimensions similar to sensitive nanometric DNA targets.
Particularly important are the nanodosimetric quantities $F_k$ ($k=1,2,3$), which indicate the cumulative probabilities to induce clusters of size  $\ge k$ in a nanometric volume. $F_2$ is known to be correlated to the probability of inducing DNA double strand breaks (DSB), while $F_3$ is connected to the probability of producing complex lethal damage.
 
\begin{figure}[t]
\centering	
\includegraphics[width=\linewidth]{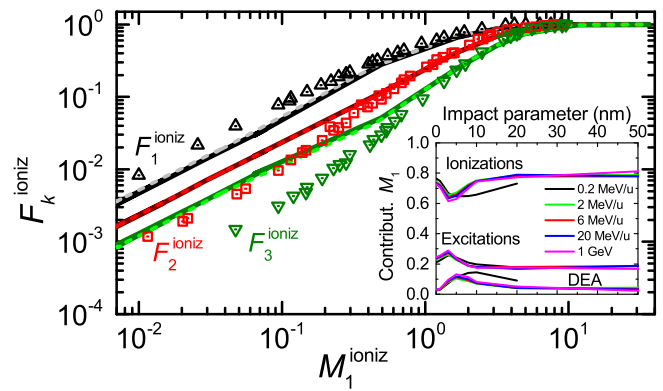}
\caption{Simulation results of the ionization events $F_k^{\rm ioniz}$ ($k=1,2,3)$ vs. $M_1^{\rm ioniz}$, corresponding to several energies of the incident carbon ion and different impact parameters. Solid (dashed) lines correspond to a nanometric volume of 20 (10) base pairs. Symbols constitute a compilation of experimental data \cite{Conte2018}. The inset shows the fractional contribution to the average cluster size due to ionization, excitation (with only $40\%$ of them leading to molecular dissociation) and DEA events, at several carbon ion kinetic energies and impact parameters.
}
\label{fig:ClusterDistribution}
\end{figure}

Remarkably, it is known that the representation of the measured $F_k^{\rm ioniz}$ distributions for ionization events as a function of the average ionization cluster size $M_1^{\rm ioniz}$ yields a universal distribution, independent of the size and characteristics of the particular nanodosimeter, which can be used to predict cell inactivation cross sections \cite{Conte2017,Conte2018}. Our simulations provide these ionization distributions in nanometric cylinders of liquid water, mimicking DNA targets, at different values of the carbon ion $T$ and $r$. Figure \ref{fig:ClusterDistribution} shows $F_k^{\rm ioniz}$ ($k=1,2,3$) for ionization in the range 0.2 MeV/u--1 GeV and $0 \mbox{ nm} \le r \le 100 \mbox{ nm}$, as a function of the average cluster size $M_1^{\rm ioniz}$. Results are reported for two target sizes relevant for evaluating lethal DNA damage, both with diameter 2.3 nm, but having heights of 3.4 nm and 6.8 nm, respectively, corresponding to DNA turns of 10 base pairs (10 bp, dashed lines) and 20 bp (solid lines). Simulations show a good agreement with nanodosimetric measurements in gas targets in a wide range of $M_1^{\rm ioniz}$ values \cite{Conte2018}, confirming the universal relation $F_k^{\rm ioniz}$ vs. $M_1^{\rm ioniz}$ for nanometric liquid water volumes.
These data correspond to the IPs at which the largest amount of damaging events occurs, i.e. for $r < 10$ nm for every ion energy. This result confirms the accuracy of both simulations in liquid water and nanodosimetric measurements in gas targets, and also remarks that the relation $F_k^{\rm ioniz}$ vs. $M_1^{\rm ioniz}$ scales irrespectively of the size of the sensitive volume (10 bp or 20 bp DNA-like targets). Our $F_k^{\rm ioniz}$ distributions show a satisfactory qualitative behavior for the lower values of $M_1^{\rm ioniz}$ (corresponding to the higher carbon energies and/or to the larger IPs), although they deviate from experimental measurements at higher values (possibly due to differences between the gas and the liquid water targets at conditions where the number of damaging events is lower, or to the  electron-phonon and trapping cross sections). We notice that a good agreement between simulation and experimental measurements for the ionization cluster size distribution, which is the only event experimentally measured, is found.

We also use MC simulations, fed with accurate electronic excitation and ionization cross sections, as well as with the recommended DEA cross section for water \cite{Itikawa2005}, to evaluate the specific contribution of each process to the average cluster sizes, which is reported in the inset of Fig. \ref{fig:ClusterDistribution} for 0.2 MeV/u to 1 GeV carbon ion KEs and for 0 to 50 nm ion-target IPs. The relative role of the different damaging events (ionizations, excitations leading to molecular dissociation, and DEA) do not vary much neither with the carbon ion KE nor with the IP. This finding shows that ionizations contribute $\sim 80 \%$ to the average cluster size, except at $\sim 5$ nm IPs, where it is reduced to 60\%. Electronic excitations leading to dissociation, in general, add up to $\sim 20$\% of the average cluster size at all KEs, but for IPs around 5 nm it is $\sim 30 \%$. However, DEAs provide only 3\% of the average cluster size at all KEs and IPs, but only at very close IP ($\sim 5$ nm)  its relative contribution amounts to $\sim 10 \%$ of the total cluster size. For the lowest KE (0.2 MeV/u, matching the Bragg peak), an increased contribution of DEA ($15 \%$) is observed at 10 nm IP, at the expense of ionization. Note that this assessment relies on the assumption that only 40\% of excitation events lead to bond breaking. 
In light of these results we can safely state that ionization events make up the vast majority of clustered damage mechanisms in liquid water (DNA-like) targets, which supports the use of ionization-based nanodosimeters.


\section*{Conclusions}    
We have presented an accurate description of the physical processes leading to damage of living tissue (mimicked by liquid water medium) induced by swift carbon ion beams in a wide energy range addressing conditions typical of hadrontherapy and of exposure to cosmic radiation.
The main inelastic channels for charged particles, namely ionization and electronic excitation, have been obtained from accurate TDDFT first-principles calculations of the electronic excitation spectrum (energy loss function, ELF) of liquid water,
which gives access, via the dielectric formalism, to reliable inelastic cross sections to simulate the generation and propagation of SE abundantly produced by the energetic carbon ion beam passing through the liquid water medium. 
\textit{Ab initio} calculations of the elastic scattering cross section via direct solution of the Dirac equation in water clusters, to reproduce the condensed phase environment, were also carried out,
allowing precise MC simulations of the propagation of the (mostly) low energy SE and of their relevant effects at nanometer scale. 

An analysis of the clustering of damaging events in nanometric volumes mimicking sensitive DNA targets was conducted. This is important in the context of experimental nanodosimetry, relating cell response to the measurement of ionization clusters in gas-phase detectors \cite{Conte2017, Conte2018}. The nature of damaging events (measurable ionizations vs. other inelastic channels, namely dissociative excitations and DEA, not accessible to nanodosimeters) and the effects of different phases (gas vs. condensed) have been investigated with unprecedented statistical accuracy.

We find that carbon ions with energy in proximity of the Bragg peak region, such as those typically used in hadrontherapy treatments, are capable of inducing large clusters of damaging events ($>10$ for ion-target distances $\lesssim 3$ nm),
while clusters tend to be much smaller for the larger energies. 
In particular, individual high energy ions found in cosmic radiation are not capable of inducing average damage clusters larger than 1. 
We found that $\sim 70\%$ of the events leading to the damaging cluster correspond to ionization processes. The obtained ionization cumulative distributions $F_k^{\rm ioniz}$, represented as a function of the mean ionization cluster size $M_1^{\rm ioniz}$, are found in excellent agreement with nanodosimetry measurements for the larger values of $M_1^{\rm ioniz}$, where most of the biological effects occur. These findings support the use of ionization-based detectors in nanodosimetry setups. Although qualitative good agreement is found for the lower $M_1^{\rm ioniz}$ values, the presence of significant discrepancies requires deeper investigations. Nevertheless, we notice that the deviation for high-energy carbon ions may also arise owing to the phase difference used in our simulations (liquid water) and in detectors (gas-phase). 

Despite DEA being typically regarded as one of the most relevant biodamage mechanisms in radiotherapy, we have found that our simulations point towards a limited role ($\sim 10$--$15$\%) of DEA in carbon-ion induced biodamage. 
Taking into account that only a fraction of DEA leads to irreversible damage \cite{Kohanoff2017}, we conclude that this process plays an almost negligible role in carbon ion induced biodamage, being ionizations and excitations the more significant physical events resulting in harmful events due to carbon ion irradiation in a wide energy range, covering from those relevant to hadrontherapy up to cosmic radiation.




\section*{Methods}

\subsection*{TDDFT calculation of the ELF of liquid water}
\label{subsec:TDDFT-ELF}
Liquid water is a paradigmatic disordered system characterized by a large degree of randomness. To cut-off the prohibitively expensive computational task of realizing an ensemble of statistically-independent optimized water configurations, we limit our analysis to only one snapshot by assuming that the ELF is independent of the molecular configuration, as shown in \cite{Garbuio2006} for its optical response. The Supplementary Information details the procedure to implement a water supercell with which the ELF shown in Fig. \ref{fig:ELF} is calculated as well as the methods for computing the electronic excitation spectrum.


\subsection*{Cross sections of inelastic and elastic events}
\label{subsec:CrossSections}
 
The main input to simulate electron propagation in liquid water due to the energy and momentum transferred by carbon ions is, on the one hand, the set of inelastic cross sections for generating electrons, as well as their energy and angular distributions, which can be obtained from the dielectric formalism \cite{Lindhard1954,Ritchie1959,Nikjoo-Uehara-Emfietzoglou2012} and, on the other hand, electron elastic cross section, obtained by solving the Dirac equation of the electron in the field of the target atoms \cite{morresi2018nuclear}. The procedure to obtain these cross-sections is presented in the Supplementary Information.

\subsection*{Generation, transport and effects of secondary electrons }
\label{subsec:EnergyAngularDistributionsSecondaryElectrons}

To reckon the electron transport within liquid water and the generation of the SE cascade we have used the event-by-event MC code SEED (Secondary Electron Energy Deposition) \cite{Dapor2017,Dapor2020book}, which follows electron trajectories accurately accounting for inelastic events (typically at higher energy), resulting into excitation and ionization of the medium atoms, as well as elastic collisions with atomic ions of the target (typically occurring below 50 eV), which lead to a change in trajectory. Events particularly noticeable at low electron energies, such as electron-phonon and electron-polaron loss interactions, as  well as DEA processes, are also included in the simulation, as explained in the Supplementary Information.


\section*{Acknowledgements}
MD and ST acknowledge the Bruno Kessler Foundation and the National Institute of Nuclear Physics for unlimited access to their computing facilities, and the Caritro Foundation for the grant {\it High-Z ceramic oxide nanosystems for mediated proton cancer therapy}. PdV has been founded by a Juan de la Cierva fellowship (FJCI-2017-32233). This work was also supported by the Spanish Ministerio de Ciencia e Innovaci\'on and the European Regional Development Fund (Project PGC2018-096788-B-I00), by the Fundaci\'on S\'eneca -- Agencia de Ciencia y Tecnolog\'ia de la Regi\'on de Murcia (Project 19907/GERM/15) and by the Conselleria d'Educaci\'o, Investigaci\'o, Cultura i Esport de la Generalitat Valenciana (Project AICO/2019/070).

\end{document}


\title{On the relative role of the physical mechanisms \\on complex biodamage induced by carbon irradiation \\ (Supplementary Information)}
\author{Simone Taioli$^{a,b,c}$}
\email{taioli@ectstar.eu}
\author{Paolo E. Trevisanutto$^{a,b,d}$} 
\author{Pablo de Vera$^e$}
\author{Stefano Simonucci$^{f,g}$}
\author{Isabel Abril$^h$}
\author{Rafael Garcia-Molina$^e$}
\author{Maurizio Dapor$^{a,b}$}
\email{dapor@ectstar.eu}

\affiliation{$^a$European Centre for Theoretical Studies in Nuclear Physics and Related Areas (ECT*-FBK)}
\affiliation{$^b$Trento Institute for Fundamental Physics and Applications (TIFPA-INFN), Trento, Italy}
\affiliation{$^c$Peter the Great St. Petersburg Polytechnic University, Russia}
\affiliation{$^d$Center for Information Technology, Bruno Kessler Foundation, Trento, Italy}
\affiliation{$^e$Departamento de F{\'i}sica, Centro de Investigaci{\'o}n en {\'O}ptica y Nanof{\'i}sica, Universidad de Murcia, Spain}
\affiliation{$^f$School of Science and Technology, University of Camerino, Italy}
\affiliation{$^g$INFN, Sezione di Perugia, Italy}
\affiliation{$^h$Departament de F{\'i}sica Aplicada, Universitat d'Alacant, Spain}


\maketitle


\subsection*{TDDFT calculation of the ELF of liquid water}
\label{subsec:TDDFT-ELF}

The energy loss function (ELF) of a material provides its electronic excitation spectrum in the momentum and energy space $(\hbar k, E)$. It is obtained from its complex dielectric function $\epsilon (k, E)$ as ${\rm Im} [-1/\epsilon(k,E)]$. 

A water supercell was generated by carrying out molecular dynamics (MD) simulations with several thousands molecules, using the empirical TIP3P force-field \cite{doi:10.1021/jp973084f} implemented in the LAMMPS package \cite{LAMMPS}. The simulations ran for 100 ps, the first 10 ps being due to reach thermodynamic equilibrium at $T=300$ K. 
A cubic cell with side of 0.985 nm that can accommodate 32 water molecules to reproduce the experimental water density at room conditions ($\rho=1$ g/cm$^3$) was then obtained. The latter cell size is a trade-off between reasonable computational effort of the many-body calculations and good agreement with experimental ELF data
\cite{Watanabe1997,Hayashi2000,Watanabe2000}. 
Finally, this cell was further relaxed imposing periodic boundary conditions below $10^{-3}$ Ry/$\AA$ for the interatomic forces via first-principles simulations based on density functional theory (DFT) \cite{Hohenberg-PR-136-B864}, implemented in the Quantum Espresso code suite \cite{Giannozzi_2009}.
The optimized water configuration used in this work appears in Fig. S1.

\begin{figure}[b]
	\centering
	\includegraphics[width=0.33\textwidth]{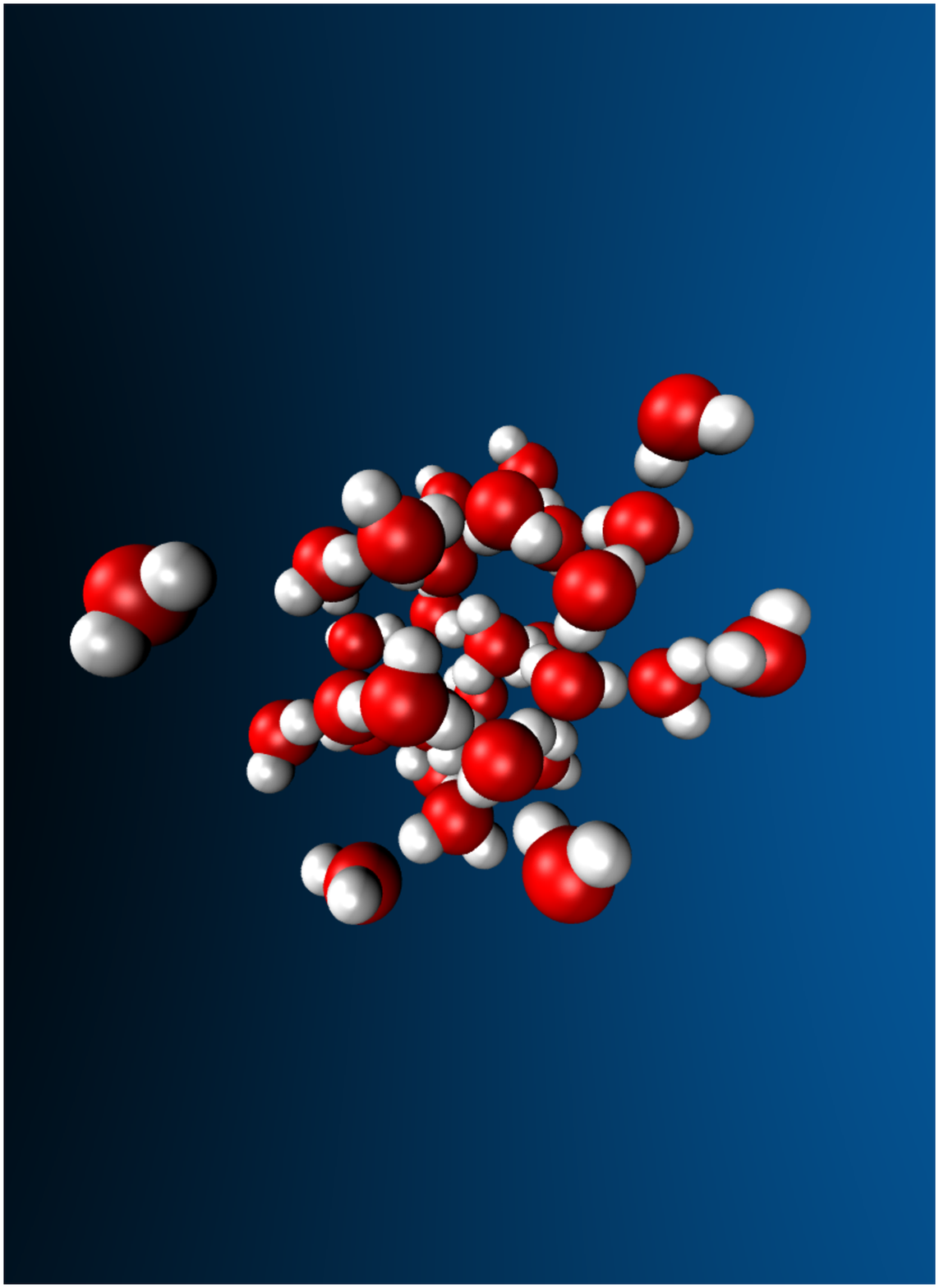}
	\caption{Optimized water configuration used to calculate the ELF via TDDFT.}
\end{figure}

The assumption that a single snapshot of the liquid water configuration is enough to obtain its ELF$(k,E)$ relies on previous photoabsorption spectra simulations of liquid water, where different molecular arrangements showed similar optical response \cite{Garbuio2006}.

Due to the random orientation of water molecules, only the dependence on the wave vector module $k$ was considered in the ab-initio calculations, performed by the LR-TDDFT approach \cite{RevModPhys.74.601} with PBE exchange-correlation kernel \cite{kernel} to include the electron-hole interactions in the ELF spectra. Lanczos chains algorithm, implemented in the turboEELS code \cite{Turbo-eels}, was used to avoid the sum over excited states.  
The water ELF converged with a $4 \times 4 \times 4$ Monkhorst-Pack mesh grid and $600$ Lanczos iterations.

\subsection*{Cross sections of inelastic and elastic events}
\label{subsec:CrossSections}

The dielectric formalism \cite{Lindhard1954,Ritchie1959,Nikjoo-Uehara-Emfietzoglou2012} provides a theoretical framework to study the inelastic interactions of charged particles with matter, in such a way that the features of the projectile (charge, mass and energy) and the medium (electronic excitation spectrum, through its ELF) appear decoupled in all the expressions used.

The basic quantity to study the generation and subsequent transport of electrons in a medium resulting from the interaction with a swift charged particle is the probability of transferring an energy $E$ and a momentum $\hbar k$ to the medium, which is provided by the doubly differential cross section DDCS \cite{Lindhard1954,Ritchie1959,Nikjoo-Uehara-Emfietzoglou2012}: 
\begin{equation}
\frac{{\rm d}^2\sigma}{{\rm d}E \, {\rm d}k} = \frac{e^2}{\pi \hbar \, {\cal N}}\frac{M [Z-\rho_q(k)]^2}{T}\frac{1}{k} \, {\rm ELF} (k,E)  \, ,
\label{eq:DDCSkw}
\end{equation}
where $Z$, $M$, and $T$ are the  atomic number, mass and energy for the case of an incident ion; the charge state $q$ of the ion is accounted for through the Fourier transform $\rho_q(k)$ of its electronic density. The response of the medium to the perturbations created by the external charged particle is provided by its energy loss function  ${\rm ELF} (k,E)$. 


The number of generated electrons, their energy and angular distributions can be obtained from the DDCS \eqref{eq:DDCSkw} for a carbon ion by integration through energy and/or momentum transfer, or a suitable transformation to obtain the dependence in ejection angle \cite{deVera2013c,deVera2015}. 
The energy $E$ delivered by the charged projectile to the electronic degrees of freedom of the medium is related to the energy $W$ of an ejected electron through $E=B+W$, with $B$ being the mean binding energy of the outer-shell electrons, which discriminates whether an excitation (if $E<B$) or an ionization (if $E>B$) occurs \cite{deVera2013c,deVera2015,deVera2019}.
\eqref{eq:DDCSkw} can be also used to determine the cross sections for ionization and excitation due to electron impact by replacing the ion characteristics by the electron ones ($M=m$, $Z-\rho_q(k)=1$) and, additionally, by suitably choosing the integration limits and accounting for exchange and indistinguishability effects \cite{deVera2019}.

The elastic cross sections for an electron moving through a medium are obtained by direct solution of the Dirac equation in a multi-centric functional space to account for the randomly oriented molecular system. In particular, wavefunctions and self-consistent potentials are expanded in a basis set of aug-cc-pVTZ Gaussian functions (GBS), centered into the nuclei. Mono-- and bi-- electronic molecular integrals are computed at each SCF cycle among the GBS functions, and then through a unitary transformation in the molecular orbital basis \cite{morresi2018nuclear}. To reduce the computational cost, a cluster of six water molecules is considered. Only the potential term in the Dirac Hamiltonian is projected onto the finite set of $L^2$ functions to recover the continuum. The multi-scattering interference terms are inherently included in the formalism. The differential elastic cross section for solid angle unit is then obtained as follows:
\begin{equation}
\frac{\mbox{d}\sigma}{\mbox{d}\Omega}=\frac{m^2}{4\pi^2}|\langle \phi_{k\hat{n}}|T^+(E)|\phi_{\bf{k}}\rangle|^2=\frac{m^2}{4\pi^2}|\langle \phi_{k\hat{n}}|V|\psi_{\bf{k}}^+(\textbf{r}')\rangle|^2 
\end{equation}
where $m$ is the electron mass, $\phi_{k\hat{n}}$ is the incoming plane-wave impinging on the water cluster with momentum $k$ in the direction $\hat{n}$, $\phi_{\bf{k}}$ is the outgoing free plane wave elastically scattered in the direction $\bf{k}$ within the solid angle $\Omega$ and $(\Omega+\mbox{d}\Omega)$, $T^+(E)$ is the on-shell $T$-matrix, $V$ is the self-consistent molecular potential obtained by the solution of the Dirac equation, and $\psi_\textbf{k}^+(\textbf{r}') = \exp( \mbox{i} \textbf{kr})-\frac{m}{2\pi}\frac{\exp(\mathrm{i}\textbf{kr})}{r}\psi_\textbf{k}^+(\textbf{r}')$ is the scattering wavefunction.
Since $V$ is the approximate representation of the long range Coulomb potential projected on a finite functional space, one can replace $\psi_{\bf{k}}^+$ with $\phi_{\bf{k}}$ outside the scattering volume where the potential dies off \cite{taioli2009surprises,Taioli2010}.

\subsection*{Generation, transport and effects of secondary electrons }
\label{subsec:EnergyAngularDistributionsSecondaryElectrons}

Monte Carlo simulations at each carbon kinetic energy $T$ were carried out using 1200 ion's tracks of 50 nm length each one, with different random seeds at each ion's shot. The latter path length was chosen so that virtually all the secondary electrons  generated along the carbon ion track reach the sensitive volume (having dimensions of a DNA-like target), while keeping simulation times within reasonable limits. To achieve an acceptable trade-off between computational cost and low signal-to-noise ratio, we assume that 1000 electrons are generated initially along the track at each collision between the carbon ion and the water target. In average, carbon ions undergo 30 (1 GeV) to 1000 (0.2 MeV/u) collisions; thus each ion shot produces on average $10^5$--$10^6$ electrons. These electrons then scatter within the target material, producing an average number of 100 further electrons each, before stopping; multiplying by the number of ion shots (1200) our simulations are equivalent to assess 4 to 100 billion electron trajectories, which were followed up by means of the Monte Carlo code SEED (Secondary Electron Energy Deposition) \cite{Dapor2017,Dapor2020book} 
until absorption in the medium. Deviation of the electron trajectory was accounted for through the elastic cross section, and different inelastic events (ionization, excitation, DEA, electron-phonon and electron-polaron) were drawn according to their relative probability$p_i =(\sum_i \Lambda_i)/\Lambda_i$, where $\Lambda_i$ is the inverse mean free path between two collisional events of the $i$ type, using a Bortz, Kalos and Lebowitz (BKL) acceptance algorithm \cite{growcopp,Bortz197510}. The probability of having a scattering process $i$, be it elastic or inelastic, is compared with a random number and the type of collision is selected. Depending on the event, the electron trajectory and energy are modified. To determine the nanodosimetric observables presented in this work, the possible damaging events (ionization, excitation or DEA) are scored only when occurring inside the sensitive volume for each distance from the ion's track.

\bibliography{SIbib}









